\RequirePackage{amsthm}
\documentclass[default,iicol]{sn-jnl}

\usepackage{graphicx}%
\usepackage{multirow}%
\usepackage{amsmath,amssymb,amsfonts}%
\usepackage{amsthm}%
\usepackage{mathrsfs}%
\usepackage[title]{appendix}%
\usepackage{xcolor}%
\usepackage{textcomp}%
\usepackage{manyfoot}%
\usepackage{booktabs}%
\usepackage{algorithm}%
\usepackage{algorithmicx}%
\usepackage{algpseudocode}%
\usepackage{listings}%
\usepackage{cellspace} 
\usepackage{capt-of}

\RequirePackage{fix-cm}

\newcommand{\eg}{\textit{e.g.}, }
\newcommand{\apo}{\textit{a posteriori} }
\newcommand{\apr}{\textit{a priori} }

\usepackage{amsmath,amssymb}

\usepackage{ulem}
\usepackage{xcolor}
 
 \newcommand{\Autoref}[1]{%
  \begingroup%
  \def\chapterautorefname{Chapter}%
  \def\sectionautorefname{Section}%
  \def\subsectionautorefname{Subsection}%
  \autoref{#1}%
  \endgroup%
}

 \usepackage{ulem}
 \newcommand{\pvec}[1]{\vec{#1}\mkern2mu\vphantom{#1}}

\begin{document}
\title{Parameter optimisation using Bayesian inference for spallation models}

 \author*[1,2]{J.~Hirtz}\email{jason.hirtz@unibe.ch}  
 \author[1]{J.-C.~David}  
 \author[3]{J.~Cugnon}  
 \author[2]{I.~Leya}  
 \author[4]{J.L.~Rodríguez-Sánchez}  
 \author[5]{G.~Schnabel}  

 \affil[1]{IRFU, CEA, Universit\'{e} Paris-Saclay, F-91191, Gif-sur-Yvette, France}
 \affil[2]{Space Research and Planetary Sciences, Physics Institute, University of Bern, Sidlerstrasse 5, 3012 Bern, Switzerland}
 \affil[3]{AGO department, University of Li\`ege, all\'ee du 6 ao\^ut 19, b\^atiment B5, B-4000 Li\`ege, Belgium}
 \affil[4]{CITENI, Campus Industrial de Ferrol, Universidade da Coru\~{n}a, E-15403 Ferrol, Spain}
 \affil[5]{NAPC-Nuclear Data Section, International Atomic Energy Agency, Vienna, Austria}

\abstract{
 The accuracy and precision of high-energy spallation models are key issues for the design and development of new applications and experiments.
 We present a method to estimate model parameters and associated uncertainties by leveraging the Bayesian version of the Generalised Least Squares method, which enables us to incorporate prior knowledge on the parameter values.
 This approach is designed to adjust parameters based on experimental data, accounting for experimental uncertainty information, and providing uncertainties for all adjusted parameters.
 This approach is designed in order both to improve the accuracy of models through the modification of free parameters of these models, which results in a better reproduction of experimental data, and to estimate the uncertainties of these parameters and, by extension, their impacts on the model output.
 We aim at demonstrating the Generalised Least Square method can be applied in the case of Monte Carlo models.
 We present a proof-of-concept for Monte Carlo models in the specific case of nuclear physics with the model combination INCL/ABLA.
 We discuss the challenges in the application of this method to high-energy spallation models, notably the large runtime and the stochasticity of the models.
 Our results indicate this framework can also be applied to analogous situations where parameters of a computationally expensive Monte Carlo code should be inferred/improved.
 }

\maketitle

\section{Introduction}
\label{intro}

 As the Dutch physicist Walter Lewin wisely said: ``Any measurement that you make, without any knowledge of the uncertainty, is meaningless''.
 It is true for experimental measurements as well as for theoretical models.
 As precise and reliable as they can be, experimental data and models are always only an approximation of the reality and, therefore, the difference between data or models predictions on the one hand and the reality on the other hand have to be estimated to make them meaningful.
 However, while the estimation of uncertainties for experimental measurements became the norm a century ago, the evaluation of model uncertainties is much more recent and was a long time limited to the consideration of statistical uncertainties only.
 
 Nuclear physics found a wide range of application (\eg fusion technology, medical hadron therapy, cosmogenic nuclide production, transmutation of nuclear waste, etc.).
 The study of the nuclear data used, means and uncertainties, is commonly called \textit{nuclear data evaluation} and is a critical aspect for these applications.
 However, measuring all required nuclear data is impossible for all the various fields of application.
 Models able to predict the relevant data are of the highest importance in the field of nuclear physics as they are needed to design instruments, for radioprotection, or simply to analyse experimental data.
 The improvement of computing power in the last decades allowed the development of new tools for model uncertainty quantification, especially for Monte Carlo (MC) models.
 Considering the various fields of applications of nuclear models and their relevance for societies, it is obvious that model calculations must be as precise and reliable as possible.
 Consequently, the bias, \textit{i.e.}, the difference between the estimator and the true value of an observable, and the uncertainties of models must be estimated precisely for a proper use of these nuclear models.
 
 In the past decades, various methods have been developed to estimate model parameters and associated uncertainties.
 Many of them are based on Bayesian statistics.
 One can mention various approach in the nuclear data field \citep{tmc,talys,bmc1,bmc2,fmc,bfmc,umc1,umc2,tmcumc} and Bayesian inference for R-matrix fitting \citep{desouza,desouzaAnswer}.
 More recently, evaluation approaches explore and employ Bayesian hierarchical modelling, \eg \citep{georg2, Mumpower}.
 Bayesian methods have also been studied and employed in the wider nuclear physics field, \eg \cite{band}.
 We may also mention the early seminal work of Kennedy and O'Hagan \cite{Kennedy} demonstrating the Bayesian approach for uncertainty quantification of expensive black-box computer models.
 
 Bayesian statistics is a general framework for inference where limited knowledge about quantities is expressed in terms of probability distributions, see \eg \cite{tuto} for an introduction.
 The object of central interest in Bayesian inference is the posterior distribution, which represents an updated state of knowledge taking into account observations (entering the likelihood) and prior knowledge.
 This allows us to estimate the likelihood of a result as well as its uncertainties.

 In the 20th century, nuclear data evaluation was mostly focussed on neutron-induced reaction with energies below 20 MeV.
 This led to the creation of nuclear data libraries \citep{jendl, endf, jeff}, which are tables of nuclear-physics observables needed for application simulations.
 At present, new types of projects are envisaged with much higher operating energies and with more types of projectile particles.
 As an example, the \textit{Multi-purpose hYbrid Research Reactor for High-tech Applications} (MYRRHA \cite{myrrha}) project will operate at energies up to 600~MeV.
 Therefore, a new (and large) energy range must be carefully studied.
  
 During the European Nuclear Data project ``solving CHAllenges in Nuclear DAta'' (CHANDA) \cite{chanda}, and more recently in the ``Supplying Accurate Nuclear Data for energy and non-energy Applications'' (SANDA) \cite{sanda} project, an important effort has been devoted to the development, improvement, and validation of high energy nuclear models, in particular the combination of the IntraNuclear Cascade model of Liège (INCL) \citep{inclBoudard,inclDavide,bibi} and the Ablation model (ABLA)\citep{JL2022, JL2023} that are now widely used for high energy applications.
 
 INCL is a MC model devoted to the simulation of spallation reaction: the interaction of light particles (proton, neutron, light cluster, etc.) with heavier target nucleus within the energy range from few tens of MeV to a few GeV.
 Initially, the target nucleus is described as a Fermi gas.
 The projectile is then shot in direction of the target which might result in a collision.
 The entering nucleons will result in an intranuclear cascade of binary collisions between the hadrons present.
 When a particle from this cascade reaches the surface of the nucleus, it has the possibility to be emitted depending on its energy.
 Its main ingredients, which can be modified in order to improved the model prediction, are the binary cross sections, features describing the initial state of the target nucleus (\eg Fermi momentum), and particles properties (\eg Pauli blocking parameters).
 To be complete, INCL is often associated to the ABLA model, which is able to simulate the de-excitation of the remnant or compound nucleus obtained at the end of the cascade.
 ABLA treats the de-excitation through different processes in competition.
 Namely, the evaporation of light particle (\eg $\gamma$, p, $\alpha$) using the Weisskopf-Ewing theory, the Fission, and the multi-fragmentation, also called Fermi break-up.
 The model requires various parameters like the fission dissipation coefficient or emission barrier corrections for light ions.
 The combination INCL/ABLA has been evaluated as the best between all the configuration tested \cite{iaea}.

 In the CHANDA project \cite{chanda}, for the first time, a study had been conducted to investigate a possible methodology based on the Bayesian framework for quantifying the uncertainties linked to parameters in high energy models, which could then possibly be taken into account in MC transport codes \cite{georgChanda}. 
 In the present study, which was included in the SANDA project \cite{sanda}, it is proposed to investigate if the methodology we developed can be applied to a large number of parameters used in the INCL model and if the methodology can be applied within a reasonable computational time.
 
 Noteworthy, the objectives of this study (see \autoref{objective}) are specific to nuclear data evaluation using the combination INCL/ABLA.
 However, the methodology developed to study our specific case is a general framework that can be applied to a large variety of models.
 In \autoref{metho}, the basics of the method is discussed together with the requirements and the limits of our approach.
 \Autoref{care} presents the treatment of experimental data required before the use of the algorithm.
 Next, the methodology is applied to INCL/ABLA with the use of real experimental data in \autoref{opti}.
 Finally, we discuss the outlook of this work in \autoref{conc}.
 
\section{Objective}
\label{objective}
 
 In the framework of the European project SANDA \cite{sanda}, we developed a method able (1) to estimate the optimal parameters for a model and (2) to estimate the uncertainties of these parameters.
 In this study, our objectives are twofold.
 First, we aim at demonstrating the feasibility of our approach for real cases using the combination of MC models INCL (for the simulation of the intranuclear cascade) and ABLA (for the simulation of the de-excitation of nuclei).
 Second, we study the possibilities, the difficulties, and the limits of our procedure both for the evaluation of the optimal parameters of a model and for the evaluation of the corresponding uncertainties.
 
 Model bias is, by definition, the expected difference between model predictions and the true values of the corresponding observables (\eg neutron multiplicity, angular distribution, mass distribution, etc.).
 Equally, the bias of the model parameters is the expected difference between the parameter values provided to the model and their true values.
 However, the ``true'' values of the parameters (when it is meaningful) are not known and are not accessible.
 Therefore, we have to rely on experimental data to characterise the model bias and the parameter bias, since this is the closest to reality once all uncertainties have been taken into account.
 
 One conceptual issue for determining parameter bias is that the definition of the bias is meaningless when the parameters are not ``physical'' parameters.
 As an example, particles masses are ``physical'' parameters, while parameters used in INCL to determine when the model stops running are model dependent parameters.
 Additionally, estimating parameter bias will be done within the Bayesian framework, which assumes that the combination INCL/ABLA is a perfect model.
 In other words, the Bayesian procedure assumes that the ``correct'' choice of parameter values will lead to predictions that perfectly coincide with the true values.
 However, as says the famous quote attributed to the British statistician George E.P. Box: ``All models are wrong, but some are useful''.
 INCL/ABLA, as any model, cannot be perfect, even with the ``correct'' parameters.
 This is why the procedure will not search for the true value of the model parameters but for the optimal parameters within the context of the model considered and of the observables studied.
 Additionally, the imperfection of our model may lead to unreasonable values for certain parameters with respect to our \apr knowledge.
 Such a case can be interpreted as a missing mechanism, an incorrect hypothesis, or a constraint not properly taken into account in the model and therefore might be used to improve the physics of the model.
 In the context of nuclear data, the inability to perfectly reproduce trustworthy experimental data is commonly refereed to as model deficiency, and different approaches have been explored to account for it, \eg \citep{leeb, Neudecker2013, georg3, Arbanas, Helgesson2018}.
 
 On the other hand, the uncertainties of the model parameters will also be evaluated.
 These uncertainties are useful as they provide information about the error propagation in the model.
 A strongly reduced uncertainty for a given parameter with respect to its \apr uncertainty would indicate that a small modification significantly modifies model predictions.
 Reciprocally, unchanged uncertainties would indicate that the model outcome is not sensitive to the exact choice of this parameter.
 Additionally, parameter correlations obtained can help understanding their relations within the model.

\section{Methodology}
\label{metho}
 
 As mentioned in \autoref{objective}, the objectives of the method we developed is to estimate the optimal model parameters and the uncertainties associated to these parameters, the latter would provide information about the error propagation in the model.
 This can be used to improve the prediction of the model both directly through the use of improved parameters and indirectly by helping model developers to find missing/badly implemented features.
 The related question of the model defects and of the estimation of model uncertainties is orthogonal to this study and has already been addressed in a previous study carried out by Schnabel within the CHANDA framework \cite{georg} and will therefore not be discussed any further.
 However, these complementary questions must be both addressed for a complete study.
 
 Our approach is divided into two main parts.
 
 In a first step, we want to know what are the optimal parameters for the model, \textit{i.e.}, we want to estimate what are the parameters that will result in the best model predictions (\textit{i.e.}, the parameter set that will maximise the likelihood of the model).
 The methodology we developed is based on the Generalised Least Squares (GLS) method \cite{gls}, which is an important technique in nuclear data evaluation.
 The GLS is often used to estimate the unknown parameters in a linear regression model, which takes into account the correlations between observed data.
 It is a method of regression similar to the common $\chi^2$ method but the correlations are taken into account, as well as the \apr values for the model parameters.
 The GLS method used here takes into account both, the reproduction of the experimental data and the \apr knowledge about the parameters, which are treated as extra data and therefore limits the risk of unphysical predictions for the parameters.
 We employ the GLS iteratively in order to account for the non-linearity of the model.
 Below, we will call this first step the GLS phase.
 
 In the second step, we want to know what are the uncertainties associated with each parameter.
 The model being non-linear, the parameters posterior distribution is not a multivariate normal distribution and cannot be directly obtained form the posterior covariance matrix obtained with the GLS method.
 Therefore, we developed an approach that can be regarded as an approximation to the Gibbs sampling \cite{Gelfand}, an iterative algorithm which will evaluate the posterior distribution and which is suitable for stochastic models.
 With this approach, we alternate between an evaluation of the posterior covariance matrix for a given parameter set using GLS formulae and a sampling of a parameter set using the posterior covariance matrix.
 The distribution of the parameters sampled along the second step allows us to determine the posterior distribution and, by extension, the uncertainties and the correlations of these parameters.
 
 It is important to mention that, if the model has difficulties reproducing some of the experimental data with respect to their error bars, the algorithm will focus on these data points and neglect others.
 This is why the selection of experimental data to be included in the analysis as well as a careful study of their uncertainties must be carried out before trying to optimise the model parameters.
 The experimental data included in this approach have to be reasonably reproducible by the model (\textit{i.e.}, all the main features involved in the corresponding process must be present in the model).
 Otherwise, these toxic data may jeopardise finding reasonable estimates for the parameter values.
 
 Rigorous uncertainty quantification of computational expensive and stochastic nuclear physics models is challenging.
 As mentioned in the introduction, numerous methods have been developed, and not only in the nuclear physics domain (\eg \citep{bio,maths}).
 For the sake of consistency with the approach that will be used by the authors in other studies to estimate model defect (Schnabel (2018) \cite{georg}),  we present a two-step uncertainty quantification procedure based on the Generalised Least Squares method and a scheme that can be regarded as an approximation to Gibbs Sampling and demonstrate its feasibility for a high-energy spallation code.
 As far as the authors know, the approximative Gibbs part is new in nuclear modelling.

 \subsection{Optimisation algorithm}
 \label{algo}
 
 In the two phases of our algorithm, an iterative algorithm is employed.
 The number of iterations for both methods is a free parameters, which needs to be specified by the user.
 For the GLS phases, it must be large enough that the approach converges to the optimal parameter set.
 For the Gibbs sampling, it must be large enough to estimate the variance of the parameters using the distribution of the parameter set produced.
 On the other hand, the computational time increases linearly with the number of iterations.
 Therefore, the minimum number of iterations required might range from a few tens to a hundred for the GLS and from a few hundreds to a few thousands for the Gibbs sampling.
 
 The main idea of the GLS is as follows.
 We start with a model $\mathcal{M}$ (here INCL/ABLA), experimental data $\vec{\sigma}_{exp}$, and a set of parameters $\vec{p}_{ref}$, which represents the best estimate of these parameters \apr (\textit{i.e.}, without knowledge of $\vec{\sigma}_{exp}$).
 Here, the model is seen as a function taking a vector as input (the parameters) and producing a vector as an output (the observables) corresponding to the experimental data.
 This means that the dimension of the model predictions, $\mathcal{M}(\vec{p})$, must be the same as the dimension of $\vec{\sigma}_{exp}$.
 In our specific case of INCL/ABLA, this is done by using an additional layer above the standard version of the model.
 This extra layer extracts the experimental setups (projectiles, targets, energies, angles, etc.) from the experimental data, runs the INCL/ABLA simulations with the same setups and with the appropriate statistics and, using the parameter set $\vec{p}$, extracts the calculated observables corresponding to the experimental data from the standard INCL/ABLA output produced and, finally, orders the observables in a vector matching $\vec{\sigma}_{exp}$.
 
 Next, we enter a loop to improve the initial set of parameters $\vec{p}_{ref}$.
 After the i-th iteration of the loop, the improved set of parameters is called $\vec{p}_{i}$.
 With the knowledge of how the model varies locally, which is given by the Jacobian (also called the sensitivity matrix) of the model evaluated at $\vec{p}_{i}$, and the difference between the model prediction $\mathcal{M}(\vec{p}_{i})$ and the experimental data $\vec{\sigma}_{exp}$, one can determine the best set of parameters $\vec{p}_{i+1}$ to minimise the difference between the model and the experimental data, assuming the model is linear between $\vec{p}_{i}$ and $\vec{p}_{i+1}$.
 Since the model is likely not strictly linear, the new set of parameters will most likely not be the optimal parameter set.
 However, as long as the model is not completely erratic between $\vec{p}_{i}$ and $\vec{p}_{i+1}$, the linearisation of the model can be seen as an acceptable approximation.
 Therefore, the new set of parameters $\vec{p}_{i+1}$ will likely be an improvement with respect to $\vec{p}_{i}$.
 Then, we can reevaluate the local Jacobian and the real model prediction in $\vec{p}_{i+1}$ and restart the loop until convergence of $\vec{p}$.
 
 Explicitly, the GLS is executed as follows.
 At the beginning of each loop, we linearise the model using a Taylor series approximation in $\vec{p}_{i}$:
 \begin{equation}
   \vec{T_i}(\vec{p}) = \mathcal{M}(\vec{p}_{i}) + J_{p_i} \times (\vec{p} - \vec{p}_{i}),
   \label{eq1}
 \end{equation}
 with the Jacobian matrix $J_{p_i}$ of the model evaluated at $\vec{p}_{i}$:
 \begin{equation}
   J_{p_i} = \left. \frac{d\mathcal{M}(\vec{p})}{\vec{dp}} \right| _{\vec{p}=\vec{p}_i}.
   \label{eq3}
 \end{equation}
 We introduce the matrix $\mathcal{J}_i$:
 \begin{equation}
   \mathcal{J}_i =
    \left(
      \begin{matrix}
         \mathbb{I}_{n\times n} & J_{p_i} \\
         \emptyset  & \mathbb{I}_{m\times m} \\
      \end{matrix}
    \right),
   \label{eq2}
 \end{equation}
 with $n$ the number of experimental data, $m$ the number of parameters, and $\mathbb{I}$ the identity matrix.
 Note that, in the case of a MC models, $\mathcal{J}_i$ is affected by the model stochasticity.
 
 The definition of $\mathcal{J}_i$ allows us to define the matrix of regression as:
 \begin{equation}
   \tilde{\Sigma}_{i} = \mathcal{J}_i \ \Sigma \ \mathcal{J}_i^T = \left(
      \begin{matrix}
         \tilde{\Sigma}_{DD_i} & \tilde{\Sigma}_{DI_i} \\
         \tilde{\Sigma}_{ID_i} & \tilde{\Sigma}_{II_i} \\
      \end{matrix}
    \right),
   \label{eq4}
 \end{equation}
 with $\tilde{\Sigma}_{DD_i}$ of dimension ${n\times n}$, $\tilde{\Sigma}_{II_i}$ of dimension ${m\times m}$, and $\Sigma$ the covariance matrix of the joint distribution of the experimental data and the input parameters:
 \begin{equation}
   \Sigma =
    \left(
      \begin{matrix}
         \Sigma_{exp} & \emptyset \\
         \emptyset  & \Sigma_p \\
      \end{matrix}
    \right),
   \label{eq5}
 \end{equation}
 with $\Sigma_{exp}$ and $\Sigma_p$ the \apr covariance matrix of the experimental data and of the parameters, respectively.
 The $\Sigma$ matrices might be non-diagonal in case of correlations between either the experimental data or the parameters.
 
 Next, we can determine an improved set of parameters using the central formula of the GLS method:
 \begin{equation}
   \pvec{p}' = \vec{p}_{ref} + \tilde{\Sigma}_{ID_i} \left( \tilde{\Sigma}_{DD_i} \right)^{-1} \left[ \vec{\sigma}_{exp} - \vec{T_i}(\vec{p}_{ref})\right].
   \label{eq6}
 \end{equation}
 The derivation of this formula is given in \autoref{preuve} for readers not familiar with the GLS method.
 
 This formula would provide directly the optimal parameters ($\vec{p}_{op}$) for the model in case the model is linear.
 However, in general, $\pvec{p}'$ is only an approximation of $\vec{p}_{op}$.
 The quality of this approximation is directly correlated to the linearity of the model between $\vec{p}_{i}$ and $\vec{p}_{op}$.
 Even if the model is not linear, $\pvec{p}'$ is likely an improvement with respect to $\vec{p}_{i}$.
 Next, we can restart the loop at \autoref{eq1} with:
 \begin{equation}
   \vec{p}_{i+1} = \pvec{p}'.
   \label{eq7}
 \end{equation}
 This will improve the quality of the GLS hypothesis of a linear model between $\vec{p}_{i}$ and $\vec{p}_{op}$ and therefore, the precision of \autoref{eq6}.
 If the model is not completely erratic, we expect the difference $|\vec{p}_{op} - \vec{p}_{i}|$ to decrease quickly with the number of iterations.
 
 For a stochastic model, the hypothesis of linearity between $\vec{p}_{i}$ and $\vec{p}_{op}$ might be reasonable as long as the expected difference of the model predictions between $\vec{p}_{i}$ and $\vec{p}_{op}$ dominates the stochasticity.
 However, as $\vec{p}_{i}$ approaches $\vec{p}_{op}$, \autoref{eq6} becomes less and less valid.
 Therefore, we expect an initial quick convergence as $\vec{p}_{i}$ is far from $\vec{p}_{op}$, then $\vec{p}_{i}$ will start oscillating around $\vec{p}_{op}$.
 In order to evaluate $\vec{p}_{op}$, we average the values of $\vec{p}_{i}$ along the oscillating phase.
 This significantly reduces the effect of stochasticity.
 
 In the second phase of the algorithm, we adopt a scheme that can be regarded as an approximation to Gibbs sampling \cite{Arbanas}.
 With the standard Gibbs sampling, we would sample alternatively between the conditional posterior distributions $\pi(\vec{p}|\mathcal{J})$ and $\pi(\mathcal{J}|\vec{p})$ and the sampled $\vec{p}$ would approximate the posterior distribution.
 
 In our scenario, we can draw samples from $\pi(\mathcal{J}|\vec{p})$ because we are able to obtain unbiased cross section predictions from our model.
 Therefore, the Jacobian estimate will then be also unbiased.
 However, we are not able to sample directly from $\pi(\vec{p}|\mathcal{J})$ because of the non-linear nature of the model which is not expressible in analytic form.
 Therefore, for each iteration, we rely on a local linearisation of the model, which would result in a multivariate normal likelihood and on the parameter prior in which one can sample.
 Then, non-linearity of the model and the stochasticity in the Jacobian estimate is accounted for through the iterations with the re-evaluation of the Jacobian.
 This approach might be a source of bias if the $\pi(\vec{p}|\mathcal{J})$ is too far from a multivariate normal distribution.
 However, for mild non-linear behaviour, the magnitude of posterior uncertainties and correlations can be evaluated with a reasonable number of iterations. 
 
 Explicitly, our approximative Gibbs sampling scheme is carried out by alternating between the evaluation of the posterior covariance matrix $\widehat{\Sigma}_{i}$ using the GLS formulae:
 \begin{equation}
   \widehat{\Sigma}_{i} = \Sigma_p - \tilde{\Sigma}_{ID_i} \left( \tilde{\Sigma}_{DD_i} \right)^{-1} \tilde{\Sigma}_{DI_i},
   \label{eq8}
 \end{equation}
 and the sampling of $\vec{p}_{i+1}$ in a multivariate normal distribution centred on $\pvec{p}'$ from \autoref{eq6} and with a covariance $\Sigma_{i}$:
 \begin{equation}
   \vec{p}_{i+1} = \mathcal{N}(\pvec{p}', \widehat{\Sigma}_{i}).
   \label{eq9}
 \end{equation}
 
 For each iteration of the Gibbs sampling, $\pvec{p}'$ is re-approximated using \autoref{eq6} (which means we process every step of the GLS but \autoref{eq7} within each iteration of the Gibbs sampling) and the covariance matrix $\widehat{\Sigma}_{i}$ is an updated version of the initial covariance matrix of the parameters (see \autoref{preuve} for details), which includes the variance of the experimental data and the error propagation through the model.
 
\begin{figure}[t]
   \centering
   \includegraphics[width=.95\columnwidth]{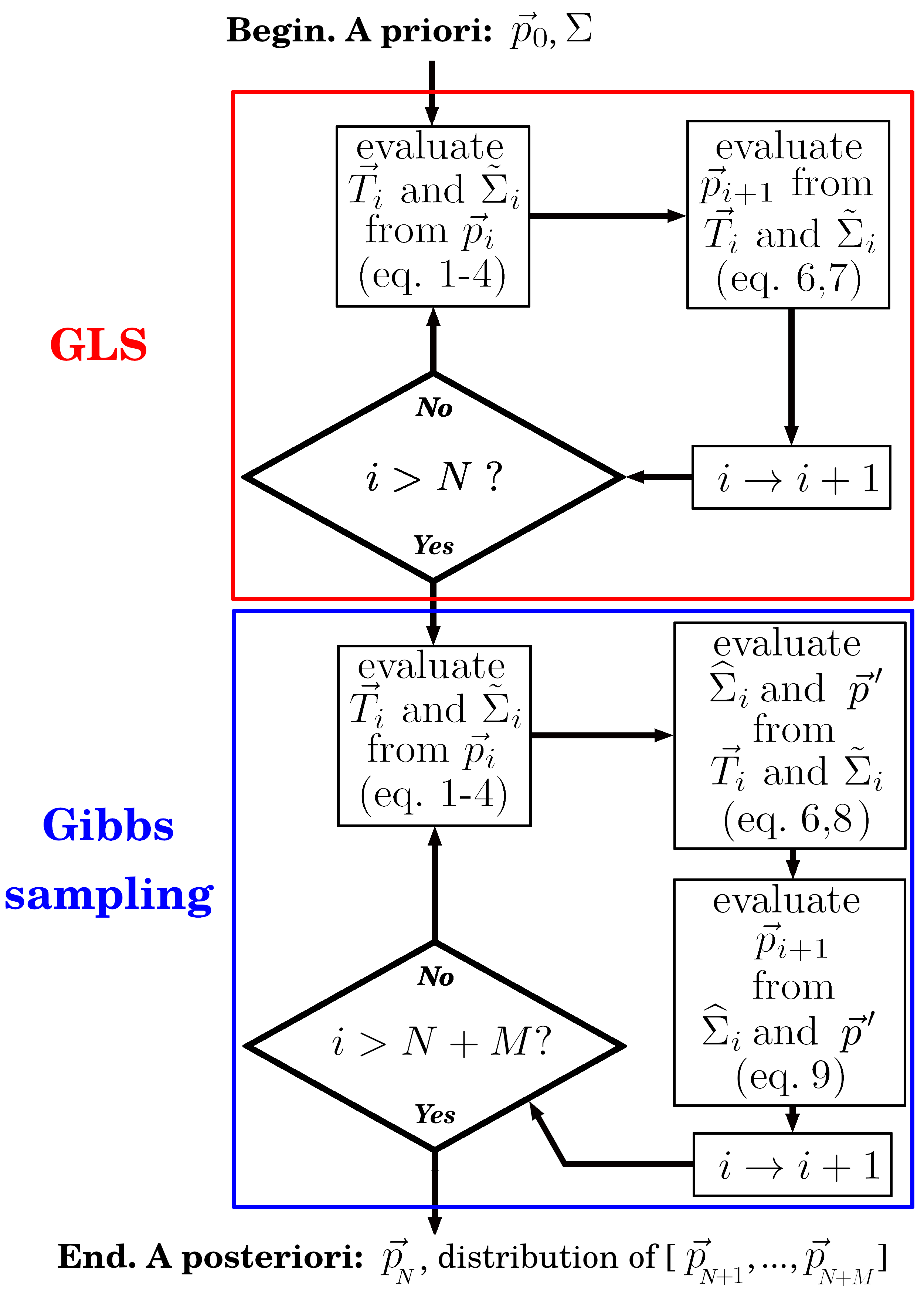}
   \caption{\label{loop} Flowchart representing the different steps of the algorithm allowing to pass from the parameter \apr $\vec{p}_0$ and the \apr covariance matrix $\Sigma$ to the posterior parameters $\vec{p}_N$ and the distribution $[\vec{p}_{N+1}$, ..., $\vec{p}_{N+M}]$.
   The latter can be turned into a posterior covariance matrix as described in the text.
   $N$ and $M$ are the numbers of iteration for the GLS and Gibbs sampling, respectively.}
 \end{figure}

 \autoref{loop} illustrates the different steps realised along the two phases of the algorithm.
 
 Note that, if we are not interested in the exact posterior distribution of $\vec{p}$, one can evaluate the posterior of the $\Sigma$ matrix directly in $\vec{p}_{op}$.
 However, this requires a statistic high enough to get rid of the model stochasticity.
 Additionally, it does not account for the non-linearity of the model, which may become noticeable if the posterior uncertainties do not constrain model parameters sufficiently for the linear approximation to hold well.
 The approximative Gibbs sampling takes into account both effects, the stochasticity and the non-linearity of the model since $\pvec{p}'$ and $\Sigma_{i}$ are re-approximated in each loop.
 This allows us to integrate the stochasticity of the model into the uncertainties of the parameters.
 The uncertainties (and the correlations between the parameters) can be extracted from the covariance matrix obtained by fitting the posterior distribution of $\vec{p}_{i}$ with a multivariate normal distribution.
 This allows us to evaluate the parameter uncertainties in the Bayesian framework taking into account the \apr uncertainties of these parameters as well as other relevant uncertainties, which are the error propagation through the model, the stochasticity of the model (which depends on the statistic used), and the uncertainties of the experimental data.
 On the other hand, if we do not want to incorporate the information about the stochasticity of the model into the uncertainties of the parameters, we can average the covariance matrix $\widehat{\Sigma}_{i}$ over the iterations.
 In this case we would obtain parameters uncertainties with only the consideration of \apr uncertainties of these parameters, \apr uncertainties of the experimental data used, and the error propagation through the model.
 \subsection{CPU Optimisation}
 
 To reduce calculation time, some CPU optimisations have been applied.
 
 First, concerning the inversion of the matrix $\tilde{\Sigma}_{DD_i}$, which is very CPU time consuming when using a large amount of experimental data, the Woodbury matrix identity is used:
 \begin{align}
   & \left( \tilde{\Sigma}_{DD_i} \right)^{-1} 
   = \left( \Sigma_{exp} + J_p \Sigma_p J_p^T \right)^{-1} \nonumber \\
   & = \Sigma_{exp}^{-1} - \Sigma_{exp}^{-1}\ J_p \ \left( \Sigma_p^{-1} + J_p^T \Sigma_{exp}^{-1} J_p \right)^{-1} J_p^T \Sigma_{exp}^{-1}.
   \label{eq10}
 \end{align}
 
 When no correlation is considered, \textit{i.e.}, when the $\Sigma$ matrices are diagonal, the inversion is straightforward.
 When there are correlations, they are often limited to a small group of data and the inversion of the matrix can still be efficiently performed.
 Consequently, the problem of a large non-diagonal matrix inversion becomes a problem of large matrix multiplications, which is faster and can easily be parallelised.
 
 Second, theoretically the Jacobian should be computed for each loop.
 However, this would be highly CPU inefficient as the Jacobian does not change drastically between two loops while the new calculation would require a significant CPU time.
 Therefore, at the end of each iteration, we check if the Jacobian is still valid and, if not, it is revaluated.
 This is done with a comparison of $\mathcal{M}(\vec{p}_{i})$ (which is evaluated in each loop for the needs of the Taylor approximation) and $T_j(\vec{p}_{i})$, the Taylor approximation performed the last time the Jacobian has been evaluated.
 If the predictions based on the exact model and the linearisation at $\vec{p}_{i}$ differ by less than a predefined value, we consider the Jacobian as still valid.
 In other words, we partially assume that the linear approximation is valid over several iterations by keeping the same Jacobian as long as this assumption is not invalidated by the computed model predictions.
 In practice, we decided to update the Jacobian only when the prediction of the Taylor approximation differs by more than twice the best relative prediction ever obtained with a Taylor approximation.
 \textit{E.g.}, if the best prediction was able to predict the model output within 20\% accuracy, we conserve the current Jacobian until the Taylor approximation differs by more than 40\% with the model output. 
 
 Third, since we assemble values expressed in different units in the $\Sigma$ matrix and, by extension, the $\tilde{\Sigma}_{i}$ matrices, it is not rare to have matrix elements that differ by many orders of magnitude.
 This can introduce errors due to the limited precision of computers while multiplying or inverting the matrices.
 In such a case, it is useful to rescale the output of the model and the experimental data.
 In other words, we can choose to optimise the parameters for the model $\mathcal{M}' = A \times \mathcal{M}$ using the experimental data $\tau' = A \times \tau$ with $A$ an arbitrary diagonal matrix.
 In such a case, the experimental error bars must be updated but not the parameters and their uncertainties.
 Proceeding this way is perfectly equivalent to optimising the parameters for the model $\mathcal{M}$ using the experimental data $\tau$.
 
 
 \subsection{Limits of the approach}
 \label{limit}
 
 In our case, in which we use INCL/ABLA, there are three main limits for the use of the method.
 
 First, one of the main challenges with nuclear data evaluation is the large number of observables to reproduce.
 One crucial assumption concerns the uncertainties of the experimental data.
 Including automatically a large number of experimental data sets into the Bayesian procedure always bears the risk that some data sets have too low uncertainties assigned.
 It is often the case with old experimental data for which systematic errors were often not evaluated or roughly set to 10\%.
 As an example, some of the experimental data included in our study had relative uncertainties below 1\%.
 In this case, the Bayesian procedure attributes a very high importance to this data while other data measured in experiments with a more rigorous uncertainty evaluation will contribute less than they should to the final results.
 Therefore, a careful study of the experimental data that are included in the Bayesian procedure must be carried out in order to use realistic (or, at least, consistent) uncertainties for every set of data. This is further discussed in \autoref{care}.
 
 Similarly, a large number of data for some experiments will lead to over fitting these data, because each data point is considered individually and not as a set of data.
 This is because most data sets almost never provide correlations.
 In other words, the more data points an experiment has, the more it will influence the final result of the study.
 For example, the neutron production cross section has been much more intensively studied than the proton production cross section.
 Therefore, if the two data sets are included in the same study, the neutron production cross sections will have much more weight for the final results than the proton production cross sections, simply because there are much more data of the former than of the latter.
 A possibility to avoid this issue, but which is out of our scope, would be to provide correlations between the data based on the related publications and/or on templates \cite{template}.
 
 The second main issue is about the stochasticity of the model used.
 The energy considered (above 20~MeV) is not described properly with deterministic models as the number of possibilities increases exponentially with energy.
 MC models become necessary for these energies but it comes with the usual balance between precision and computation time.
 However, no matter how good the statistics is, two simulations with the same initial state but different random seeds will give different results.
 In order to avoid that the \apo probability associated to a parameter set varies too much from one run to another, the statistics must be carefully chosen in order to obtain a good balance between CPU time and precision.
 This might become complex for cases with a large number of different experimental data requiring very different statistics to be properly estimated by the model.
 
 Finally, model deficiencies are not taken into account.
 Parameters can be optimised within the context of the model but the approach does not provide direct information about model deficiencies.
 An alternative approach to address the model deficiencies has been proposed by Helgesson et al. \cite{tmcumc} in which the parameter set used depends on the input.
 As mentioned in \autoref{metho}, the question of model deficiencies has been addressed in a previous work \cite{georg} in which we developed a method able to estimate the model bias.
 We decided to separate the two methods in order to focus on the physical meaning of $\vec{p}_{op}$ for the model and on the strength and limits of the approach presented here alone.
 See \autoref{favour} for an example of interpretation of $\vec{p}_{op}$ with it possible implication for future model developments.
 This has two consequences:
 First, if the model deficiencies forbid to reproduce the experimental data whatever the parameter set used, the estimated optimal set will be unsatisfactory.
 As an example, if we try to optimise the parameters of a toy model in which the data to be reproduced are distributed as a quadratic function and the toy model allows only linear functions, the approach will optimise the parameter to minimise the bias but, despite the parameter being optimal, the model will not be able to reproduce the quadratic shape of experimental data (see ref.~\cite{georgChanda}, section 3.2).
 Second, as the approach minimises the variance, which evolves with the square of the difference between experimental data and model predictions, a minor improvement in a region where the model is highly deficient will be seen as a great improvement, while a large increase of the difference between experimental data and the model predictions in regions where the model reproduce the experimental data properly will only be seen as a minor deterioration of the model.
 To summarise, if we try to optimise the model using experimental data with parts of them in deficient regions of the model due to missing/ badly implemented features and another part in efficient regions of the model, the algorithm will primarily improve the model prediction in the worst regions regardless of the effects that this produces on the model predictions in the good regions.
 In order to be complete, the quantification of the reliability of the model hypotheses should be done in a ``global'' study, \textit{i.e.}, by accounting for all the available data for which a given parameter plays a role.
 
\section{Experimental data treatment}
 \label{care}
 
 As discussed in \autoref{limit}, including a large amount of experimental data coming from a large number of experiments, teams, and from different decades, is very problematic as the quality of the data sets often differs relative to each others.
 Actually, the main problem with experimental data is not with their accuracy but with their experimental error bars, which are crucial in our analysis as they define the covariance matrix $\Sigma$.
 Sometimes, these error bars are not representative of the real accuracy and precision of the experimental data.
 Additionally, the error bars were sometimes not evaluated consistently for all experimental data sets.
 Some can be pure statistical error bars, while others include systematic errors, themselves implying a more or less thorough analysis of the experimental setup by the experimenters.
 An attempt to evaluate unknown source of uncertainties has been proposed by R. Capote et al. \cite{usu} but this approach can only be applied in cases where correction are judged to be relatively small.
 
 Sometimes, it is obvious that some of the given error bars are badly evaluated when two (or more) experimental data sets exclude each other by several $\sigma$.
 This issue has partly been addressed by Schnabel within the CHANDA framework giving the possibility to rescale automatically experimental error bars when several data sets are available for the same observables \cite{georg2}.
 However, there is only one set of experimental data available for most of the reactions we studied.
 Therefore, we need a more general approach for cases in which only one set of data is available for an observable.
 Unfortunately, to our knowledge, there is no mathematical approach allowing to provide systematic error bars for a set of experimental data based only on the experimental data themselves.
 
 One possibility to overcome this problem is the application of templates that contain reasonable ranges for the uncertainty components involved, e.g., ref.~\cite{template}.
 However, without the availability of such templates, the only way to provide reasonable uncertainty components is by thoroughly re-analysing the details provided in the publications of the experiments or interact with the experimenters, if possible.
 In cases in which it is not reasonable to reprocess the systematic error bars of all data sets included in our analysis, we propose here an alternative approach taking the error bars provided with the experimental data and applying a pragmatic algorithm to normalise those error bars.
 In \autoref{neutron}, we decided to use an algorithm ruling that experimental data with error bars too small to be realistic should be treated as experimental data with large uncertainties as there were badly evaluated to illustrate this possibility.
 Additionally, the confidence we have in those data decreases with the increasing unlikelihood of the error bars assigned.
 Therefore, the algorithm uses user-defined thresholds under which uncertainties are rescaled up to predefined levels.
 On the other hand, we decided to trust the realistic uncertainties provided by other experiments regardless of the differences of the uncertainty evaluation.
 
 In practice, all relative uncertainties below 1\% are considered as very unrealistic and are rescaled to 30\%, as well as the uncertainties not provided.
 Those between 1\% and 5\% are considered as unrealistic and are rescaled to 20\%.
 Relative uncertainties between 5\% and 10\% are considered as realistic but likely underestimated and are rescaled to 10\%.
 Finally, relative uncertainties above 10\% are considered as properly estimated and are taken as they are.
 Note that this approach forbids relative uncertainties of less than 10\%, which might be unfair for some experimental groups that made a lot of effort to reduce systematic errors.
 Such a rule-based approach, although with different rules, has also been proposed in ref.~\cite{tmcumc}.

 \begin{figure*}[t]
   \centering
   \includegraphics[width=1.9\columnwidth]{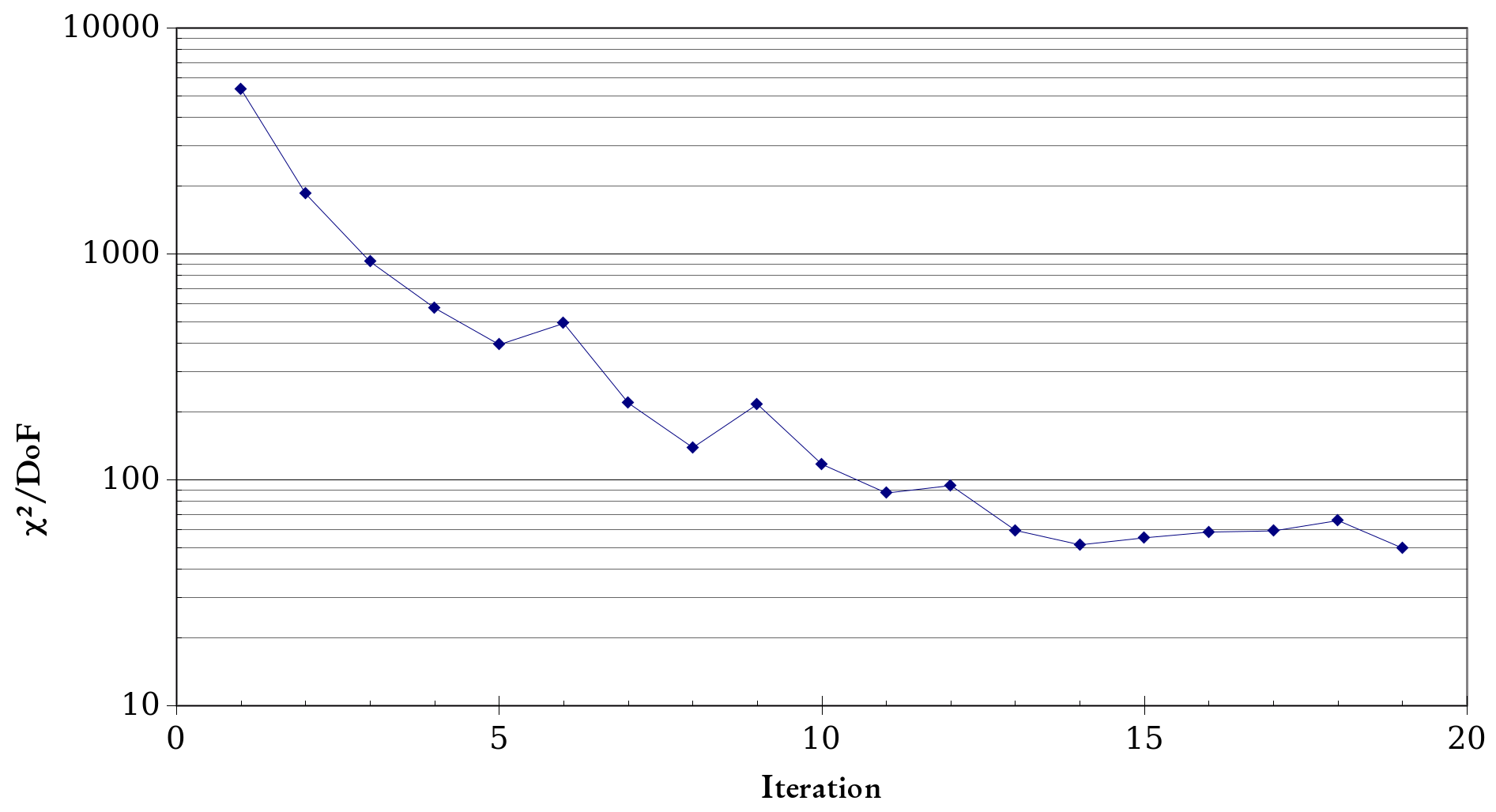}
   \caption{\label{BigImprove} Figure of merit showing the evolution of the $\chi^2/DoF$ after each iteration. Iteration 1 corresponds to the initial version of INCL.}
 \end{figure*}
 
 Such a rescaling might be needed for a proper execution of our algorithm but it has to be kept in mind that such a rule-based treatment is subjective and might have effects on the posterior.
 Although it is impossible to entirely remove the subjectivity even with more sophisticated principles or considerations, an \apo checking can be done to scale those effect, to alleviate the lack of information.

\section{Parameter optimisation}
\label{opti}
 
 When using modern models like INCL/ABLA, the parameter optimisation can be very CPU intensive, especially if ``rare'' observables are studied.
 
 Here, we study two different topics.
 First, a very favourable situation, which is not fully physically meaningful, in order to demonstrate the feasibility and the capabilities of the method.
 Second, we study a case that is representative for our long term objectives, to highlight the limits and difficulties.
 
 It is worth emphasising the two cases described below do not take into account correlations despite they are crucial to obtain meaningful results.
 See \autoref{limit} for details.
 The determination of the off-diagonal elements of the covariance matrix is a complex task, which requires both experimental and theoretical expertises.
 This exercise cannot be addressed with a simple systematic approach as described in \autoref{care} for the uncertainties.
 This aspect will be addressed in a future study.
 
\subsection{Favourable case - the subthreshold production of \texorpdfstring{$K^+$}{K+}}
\label{favour}
 
 For the first study, we chose the very favourable case of the subthreshold proton-induced $K^+$ production following the experiment at LINP \cite{LINP}.
 This case is very favourable for two reasons.
 First, the subthreshold $K^+$ production is a very specific phenomenon, which involves just a few parameters.
 Additionally, there is a limited amount of experimental data (70 data points), all coming from the same experimental set up.
 This highly simplifies both the mandatory analysis of the experimental data (see \autoref{limit}) and the analysis of the results.
 Since all data are from the same experiment, there was no need to rescale the experimental error bars as described in \autoref{care}.
 Second, the experimental data are badly reproduced by INCL \cite{bibi}, which indicates that there is large room for improvement.
 
 On the other hand, this analysis has two limitations.
 First, the phenomenon studied is a very rare event with cross sections of the order of a few nanobarns.
 Additionally, each experimental data point corresponds to a different target and different projectile energy, which requires individual calculations.
 Therefore, it is very CPU intensive to run INCL for this set-up.
 This forced us to limit the number of experimental data points used in our analysis to 24 representative points listed in \autoref{tab_data}.
 Second, the parameters involved here have an impact on other observables, which are not considered in our analysis.
 Our approach neglects the possible deterioration of such other observables that might happen when changing the parameters studied here.
 Therefore, this first study is not physically complete.
 It will be a proof of concept showing that the approach we developed is functional for complex models like INCL.
  
 \begin{figure*}[!ht]
  \centering
  \includegraphics[width=1.6\columnwidth]{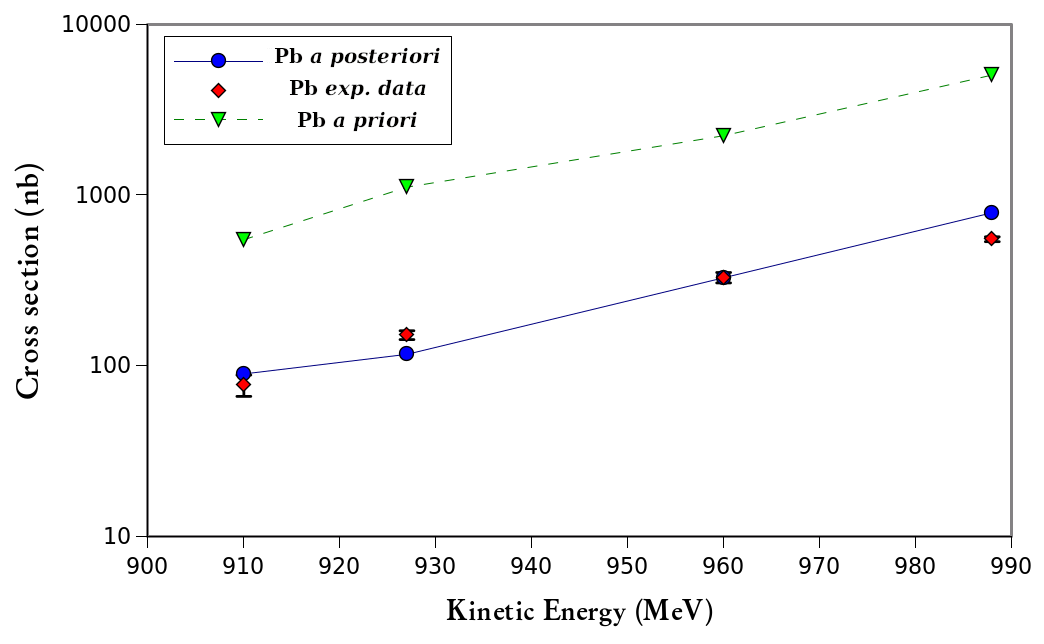}\\
  \includegraphics[width=1.6\columnwidth]{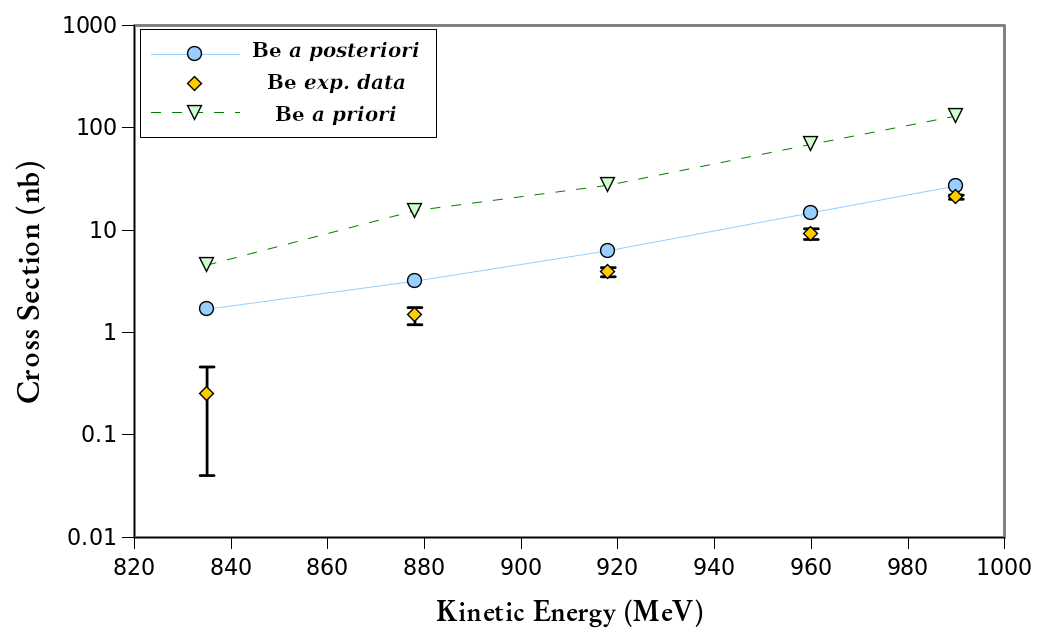}
  \caption{\label{Mieux} LINP experimental data (Diamonds) \cite{LINP} compared to INCL before (Triangles)/after (Circles) optimisation for (Top) Lead and (Bottom) Beryllium.}
 \end{figure*}
 
 We decided to consider four parameters to be optimised.
 Namely, the three scalars $a_{NN}$, $a_{\pi N}$, and $a_{\Delta N}$, which are multiplying factors applied to the original strangeness production cross sections for $NN$, $\pi N$, and $\Delta N \rightarrow K + X$ reactions, respectively ($\sigma_{x,New} = \alpha_x \sigma_{x,Old}$), and a fourth parameter, which is the Fermi momentum used in INCL.
 The \apr value for these parameters are 1.0, 1.0, 1.0, and 270~$MeV/c$ and their \apr uncertainties 0.1, 0.1, 0.1, and 5~$MeV/c$, respectively.
 No correlation was considered.
 
 \autoref{BigImprove} depicts the evolution of the $\chi^2/DoF$ after each iteration of the algorithm (see \autoref{algo}).
 Here we only used the first phase of our approach, the iterative GLS, due to CPU time restrictions.
 This calculation took 7 days using 20 cores.
 Therefore, we will not be able to provide uncertainties for the parameters.
 After only a few iterations, one can already see a huge improvement of the $\chi^2/DoF$ going from more than 5000 to roughly 50.
 The high initial value of $\sim 5300$ is explained both by the poor initial description of the experimental data (factor 5 in average) and by the rather small experimental error bars (as small as 3\%).
 Regardless of the absolute value of the $\chi^2/DoF$, the algorithm succeeds in improving the description of the experimental data by INCL as it is also illustrated in \autoref{Mieux}.
 In this figure, one can see that we started from a model highly overestimating the experimental data and we ended with a pretty fair description of the data with a factor 7.6 and 9.9 in average between the data and the model prediction before optimisation for lead and beryllium, respectively, and, after optimisation, we observe only a difference of 20\% in average for lead and a factor 2.7 for beryllium.
 
 Regarding the parameters, the algorithm multiplied the cross sections for the $NN$, $\pi N$, and $\Delta N \rightarrow K + X$ reactions by factors $1.5$, $0.26$, and $0.43$ respectively and it reduced the Fermi momentum to $232$~MeV/c.
 Theses values should not be interpreted on physical grounds, the data used being too restrictive.
 However, the study seems to indicate that there is too much energy involved in these type of reactions near the threshold and/or that the cross sections used are overestimated for the lowest energies.
 Further studies would be necessary to come to a conclusion.
 
 Overall, this example clearly demonstrates the ability of the algorithm to improve the output of a complex model like INCL through the optimisation of its parameters.

\subsection{Double differential neutron case}
\label{neutron}
 
\begin{figure*}[t]
  \centering
  \includegraphics[width=2\columnwidth]{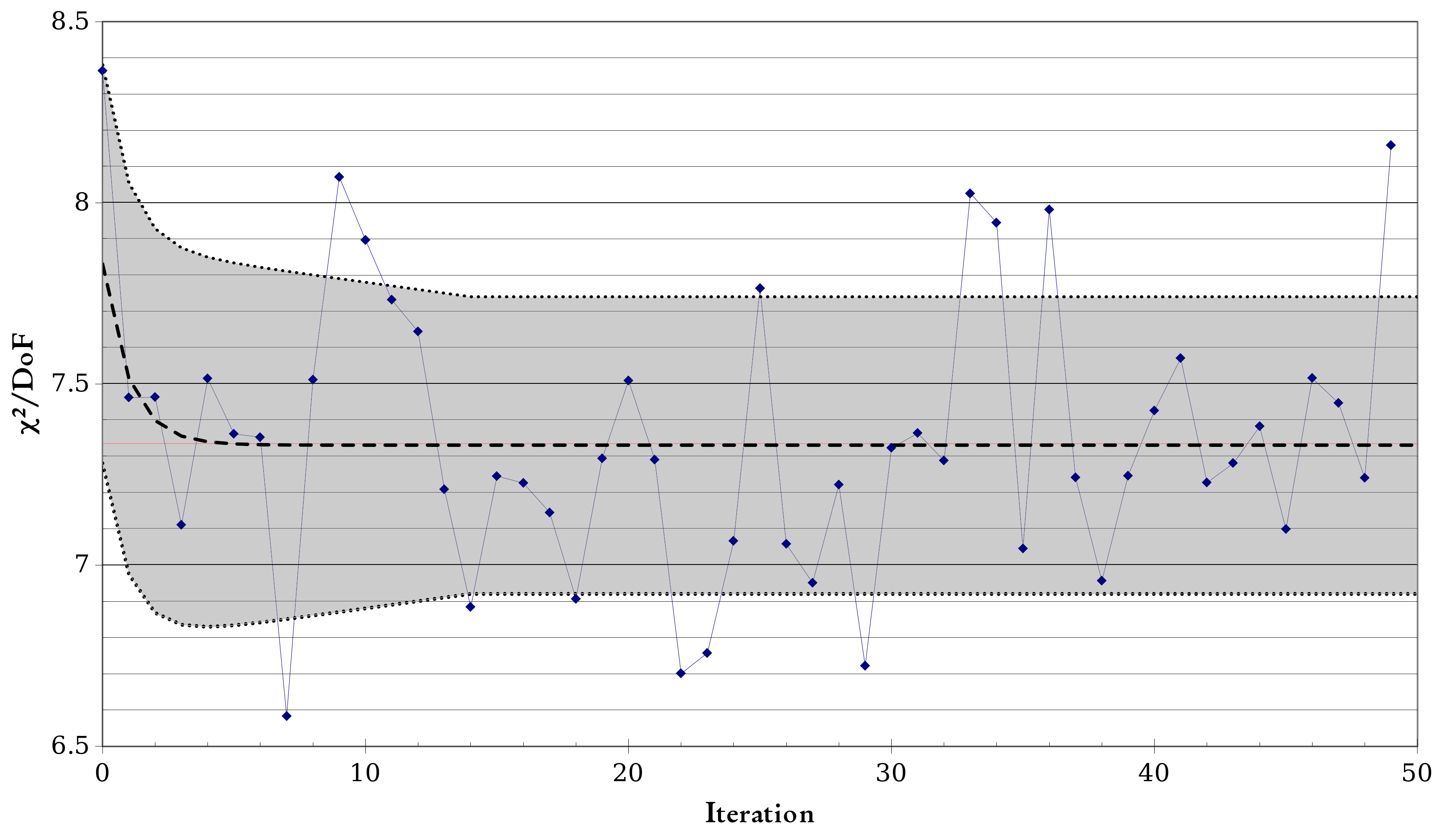}
  \caption{\label{X2evol} $\chi^2/DoF$ of INCL for the DDNXS evaluated for each iteration. The dashed line is a fit of the $\chi^2/DoF$ with the shape $a + e^{-bt}$, which is the shape expected from our algorithm. The gray area shows the 1 $\sigma$ standard deviation evaluated for the statistics used in our study.}
\end{figure*}
 
 In a second step, we decided to apply our approach to an important observable for the INCL/ABLA model applications: the double differential neutron cross section (DDNXS).
 In this case, there are much more parameters relevant for the results than in the previous example.
 One can mention almost every single elementary double differential cross section (\eg $NN\rightarrow NN$, $\pi N\rightarrow N\pi\pi$, $\Delta N\rightarrow NN$, etc.), parameters describing the structure of the nucleus, the parameters ruling clustering, the freezing-out temperature in ABLA, etc.
 Almost everything matters for such a general feature.
 Here, it is not realistic in terms of CPU power to optimise every single parameter that might be important for the DDNXS.
 It is therefore necessary to choose the parameters to be optimised.
 In our case, we have chosen to optimise a parameters scaling the $N\Delta \rightarrow NN$ cross section based on the $NN \rightarrow N\Delta$ cross section called the detailed balance parameter (DB) ($\sigma_{New} = DB \times \sigma_{detailed-balance}$), the idea originating from J. Cugnon and M.-C. Lemaire \cite{cugnon}, two parameters for the stopping time of the simulation $a$ and $b$ ($t_{stop} = a \times A^{b} fm/c$, with $A$ the mass number of the target nucleus), and the Fermi momentum.
 Based on our knowledge of the INCL model, we estimated that these parameters have enough leeway on their value, have a high impact on the DDNXS and, are therefore the most interesting to study.
 Here, we excluded parameters in ABLA to simplify our analysis.
 These parameters will be studied in future studies.
 
 Once again, because of CPU time restrictions, we limited the amount of experimental data to be taken into account.
 Here, we work with the EXFOR data base \cite{exfor} and we decided to work with proton-induced reactions with energies above 200~MeV and for target nuclei lighter than aluminium.
 We excluded experimental cross sections below $1~\mu b/sr/MeV$, which would require a very high statistics.
 This resulted in 7220 experimental data points coming from 7 publications (\citep{data1, data2, data3, data4, data5, data6, data7}).
 As mentioned in \autoref{limit}, a careful study of the experimental data used and their possible correlations must be performed in order to obtain/use the best constraints.
 The most important point in this preliminary study of the experimental data is to make sure that the experimental error bars are consistent.
 If the error bars are globally over- or underestimated, this will slightly modify the output of the optimisation, notably the error bars of the parameters, and the absolute value of the $\chi^2$.
 However, this problem is of second order compared to the problem introduced by few unrealistically small error bars aside of much more realistic but larger error bars as explained in \autoref{care}. 
 
\begin{figure*}[ht]
\centering
  \includegraphics[width=1.8\columnwidth]{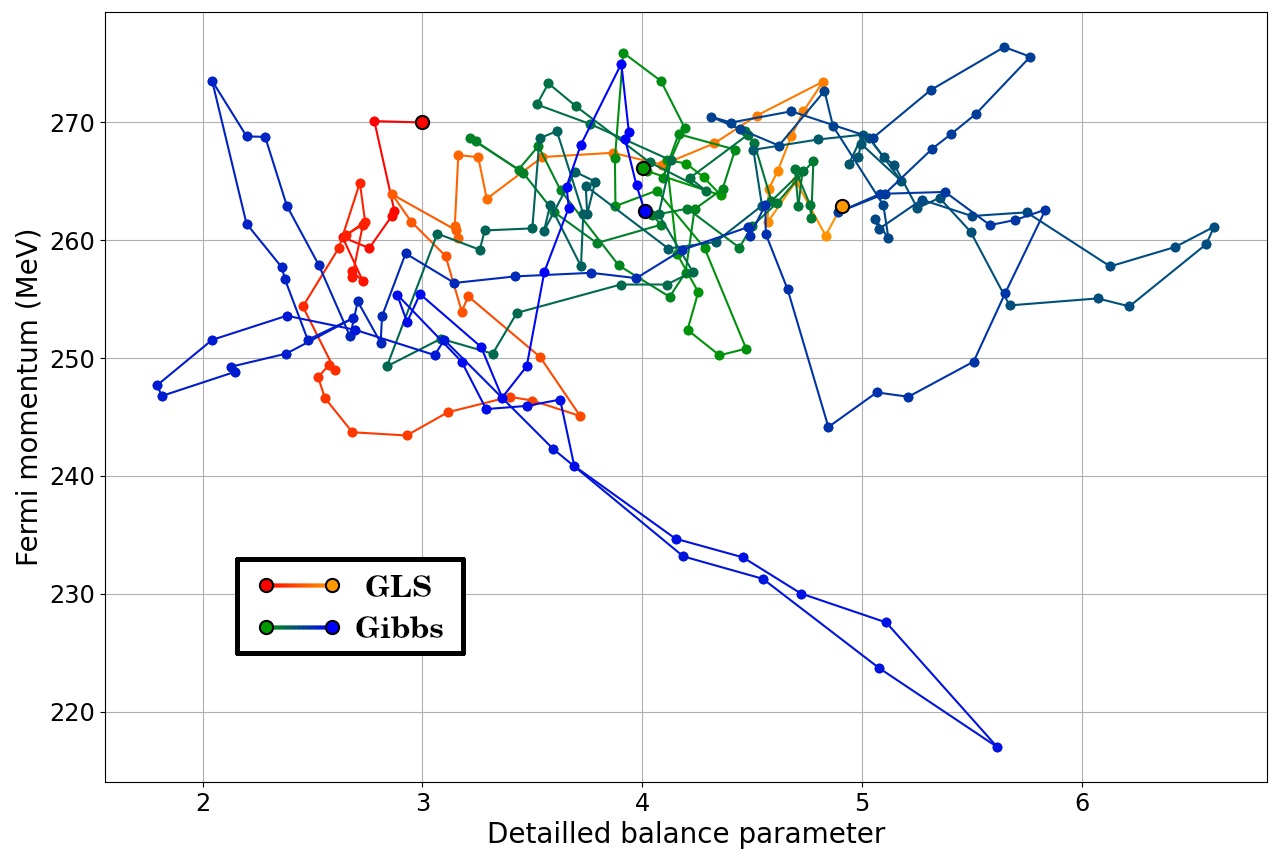}
  \caption{\label{ReTry1} Optimisation of the Fermi momentum and of the detailed balance parameters during the GLS (from red to orange) and the Gibbs sampling (from green to blue). The larger bullets indicates the initial and final values for the GLS and the Gibbs sampling.}
\end{figure*}
  
\begin{figure*}[ht]
\centering
  \includegraphics[width=1.8\columnwidth]{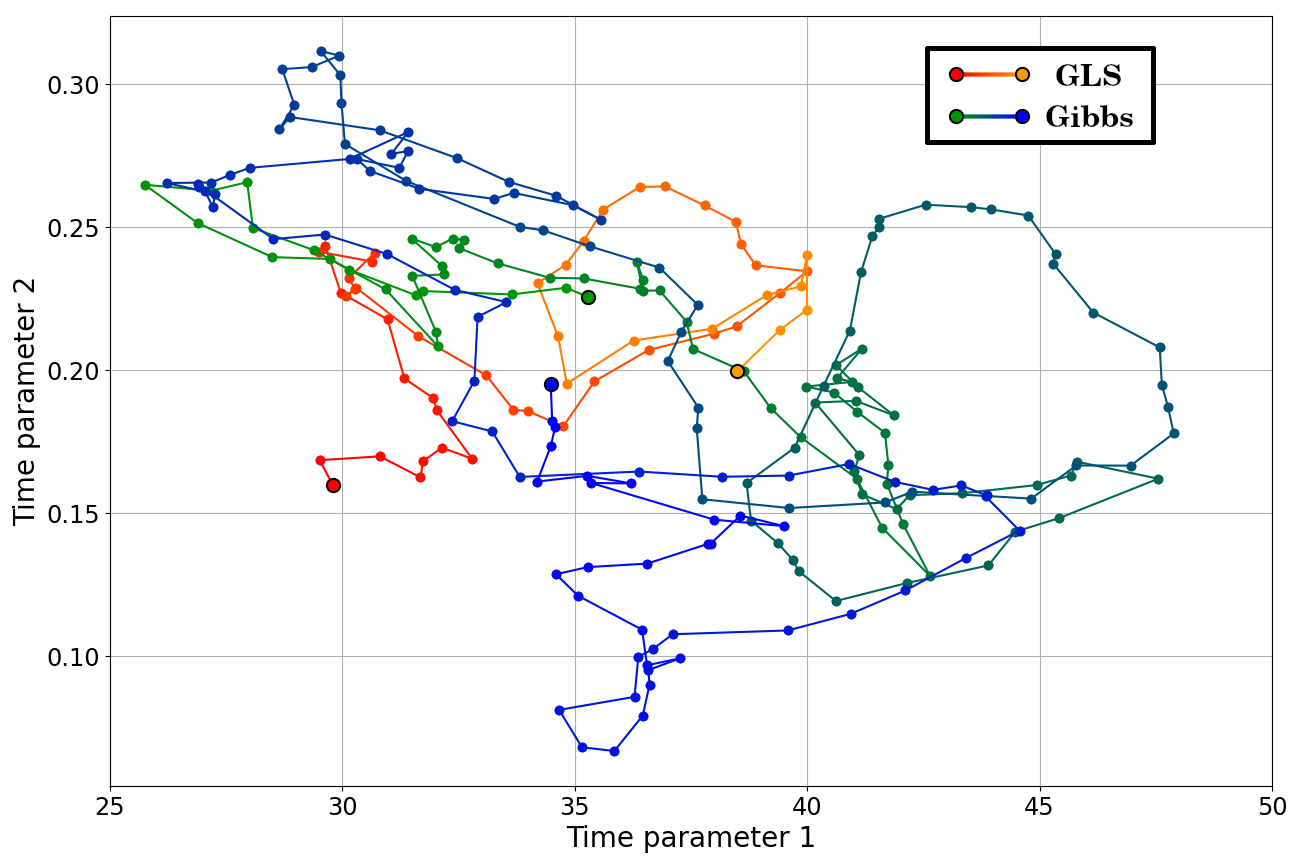}
  \caption{\label{ReTry2} Optimisation of the time parameters during the GLS (from red to orange) and the Gibbs sampling (from green to blue). The larger bullets indicates the initial and final values for the GLS and the Gibbs sampling.}
\end{figure*}

 In the case studied here, there are experimental relative error bars down to 0.45\% (EXFOR ID: C0170002, $120^\circ$ neutron emission at 2.1~MeV in the reaction p(800~MeV) + Be$^{9}$: $ 1.876 \pm 0.008406$ mb/MeV/sr).
 This kind of experimental data are toxic for our algorithm because they completely bias the value of the $\chi^2$.
 Therefore, these problematic error bars need to be rescaled.
 Otherwise, they can also be removed.
 We selected the first option.
 Our procedure to rescale experimental error bars is given in \autoref{care}.
 Our approach has not been pushed further as we are first interested in the feasibility of the method.
 
 The execution of our algorithm on the CC-IN2P3 using 20 cores took roughly 60 hours.
 
 First, we evaluated the model quality using the common reduced $\chi^2$ throughout the algorithm.
 Note, the $\chi^2/DoF$ values plotted in \autoref{X2evol} are deteriorated by various reasons.
 One can mention the quasi-elastic peak location shift between the model and the experimental data, where the uncertainties are small, the correlation missing in the covariance matrix and the fact we only use the experimental covariance matrix but not the model uncertainties:
 \begin{equation}
 \chi^2 = (\vec{\sigma}_{exp} - \mathcal{M}(\vec{p}))^T \Sigma_e^{-1} (\vec{\sigma}_{exp} - \mathcal{M}(\vec{p})).
 \end{equation}
 With this formula, a better statistics reduces the statistical uncertainties and therefore the $\chi^2$.
 The value also depends on the experimental uncertainties rescaling (see \autoref{care}).
 Using the standard values for the parameters, the $\chi^2/DoF$ is equal to $7.805 \pm 0.125$ with a standard deviation of $0.55$.
 The $\chi^2/DoF$, its standard deviation and, by extension, its uncertainty ($\pm 0.125$) was evaluated with 20 runs\footnote{$0.125 \simeq 0.55/\sqrt{20-1}$}.
 Using the optimal values as provided by our algorithm, the $\chi^2/DoF$ is now $7.34 \pm 0.094$ with a standard deviation of $0.41$.
 This represents an improvement of 6\% of the $\chi^2/DoF$.
 The two $\chi^2/DoF$ just given have been estimated with the same statistics as in the algorithm to be consistent.
 Second, the optimal parameters have been evaluated to $4.406 \pm 0.131$ (initially $3\pm 0.1$) for the detailed balance, $266.4 \pm 0.97 MeV/c$ for the Fermi momentum (initially $270\pm 3$~$MeV/c$), and for the stopping time parameters to $a = 37.13 \pm 0.59$ and $b = 0.226 \pm 0.005$ (initially $a=29.8\pm 0.5$ and $b=0.16\pm0.05$)\footnote{The uncertainties are obtained with the standard deviation of the last 20 iterations of the GLS ($ = \sigma/\sqrt{20-1}$)}.
 No \apr correlation was used even if the two time parameters should obviously be correlated.
 This is illustrated in \autoref{ReTry1} and \autoref{ReTry2} by the red/orange dots.
 The new values indicates that the cross section for the $\Delta$-recombination ($\Delta N \rightarrow NN$) has been increased by $50\%$, the maximal kinetic energy of nucleons has been slightly reduced, and the stopping time has been greatly increased ($t_{stop} = a \times A^{b} fm/c$, with $A$ the mass number of the target nucleus).
 The uncertainties are due to the stochasticity of the model, which is not fully compensated by a high number of iterations in the GLS phase of the algorithm.
 In green/blue, we show the evolution of the parameters along the Gibbs sampling.
 This provides us the range of parameter values in which the output of the model stays consistent with the experimental data.
 The \apo acceptability range for the parameters are provided by the standard deviation of the multivariate normal distribution obtained.
 Here, the 1 $\sigma$ acceptability is: $0.986$ (detailed balance), $8.822$ (Fermi momentum), $6.061$ (stopping time parameter 1), $0.0658$ (stopping time parameter 2).
 Note that the fact the initial and final values obtained are very close is a purely random effect.
 
 These uncertainties can be seen as a domain of validity given the experimental data and the model, which is considered as a valid representation of the truth.
 
\section{Summary and outlook}
\label{conc}
 
 In this study, we explored the utility of Bayesian inference, \textit{i.e.,} of the iterative Generalised Least Squares method and of an approximation of Gibbs sampling, to optimise the parameters and obtain associated uncertainties for the high-energy spallation model INCL/ABLA.
 This approach is able to fit the model predictions to the experimental data using the optimisation of the free parameters of the model.
 The objective of this algorithm is twofold:
 First, the algorithm determines optimal parameters, which minimise the bias of the model and, by extension, the $\chi^2$.
 Second, this algorithm aims at determining the uncertainties of these parameters.
 
 We demonstrated this approach based of the Generalised Least Squares method can be used for Monte Carlo models within a reasonable computing time.
 
 In our study, we first demonstrated the feasibility of the approach for a selected case, in which we reduced the $\chi^2$ of the model by a factor of 100.
 In a second stage, we studied the neutron double differential cross section with INCL/ABLA in proton-induced reactions on light target nuclei.
 We were able to produce a reasonable improvement of the model predictions by using thousands of experimental data with a reduction of the $\chi^2$ by $6\%$.
 Despite the fact that the DDNXS are extremely well studied and are already well reproduced by INCL/ABLA, thanks to our algorithm we were still able to slightly improve the model.
 Even more important, the approach is able to estimate the uncertainties of the model parameters.
 
 We also discussed the limits of the approach with, first, high CPU requirements with several days of calculation with a few tens of cores in the case of INCL.
 The application of the method will require high performance computing systems in order to reduce the time required for its execution.
 Another limit of the approach is the availability and the quality of the experimental data.
 Finally, the disparity of the quality of the experimental data is one of the most important issues, which must be addressed before applying the algorithm.
 This last point requires to exclude some questionable experimental data with unrealistically small error bars or to rescale these error bars in order to moderate their importance with respect to other data evaluated with a more rigorous approach.
 
 Once the parameters have been optimised, the model bias of the new version of INCL/ABLA can be estimated using the approach developed by Schnabel \cite{georgChanda}.
 
 Overall, these results are very encouraging showing that Bayesian methods can be used as a tool to improve the description of observables by stochastic nuclear models in the high-energy (GeV) regime.
 
\section{Acknowledgment}

 This project has received funding from the Euratom research and training programme 2014-2018 under grant agreement No 847552 (SANDA).

 NCCR PlanetS (Swiss National Science Foundation Grant Nr. 51NF40-141881) provided financial support for this research.

\begin{appendices}
\clearpage

\section{Derivation of the GLS method}
\label{preuve}

 The GLS method, \eg \cite{jeff18} section 2.2, is the basis for inference in Bayesian networks of continuous variables with a multivariate normal prior distribution and linear relationships between variables.
 Due to central importance for the study of the GLS method in general and of \autoref{eq6} and \autoref{eq8} in particular, we provide a derivation based on the Bayes' theorem.
 For a more complete description of the GLS method, the reader may consult the reference \cite{tuto}.
 
 Lets assume a set of parameters of interest $\vec{p}$, a model/function $\mathcal{M}$ that we assume to be perfect, and a set of unbiased experimental data $\vec{\sigma}_{exp}$.
 Here, a perfect model and unbiased data assumption stands for $\mathcal{M}(\vec{p}_{true}) = \vec{\sigma}_{true} = \mathbb{E}(\vec{\sigma}_{exp})$.
 
 The Bayes theorem gives the relation between the posterior distribution of $\vec{p}$ called $\pi(\vec{p}|\vec{\sigma}_{exp})$, the prior distribution of $\vec{p}$, called $\pi_0(\vec{p})$, the prior distribution of $\vec{\sigma}_{exp}$, called $\pi(\vec{\sigma}_{exp})$, and the likelihood of $\vec{\sigma}_{exp}$ knowing $\vec{p}$, called $l(\vec{\sigma}_{exp}|\vec{p})$:
 
 \begin{equation}
   \label{eqA1}
     \pi(\vec{p}|\vec{\sigma}_{exp}) = \frac{l(\vec{\sigma}_{exp}|\vec{p}) \times \pi_0(\vec{p})}{\pi(\vec{\sigma}_{exp})}.
 \end{equation}
 
 Here, $\pi(\vec{\sigma}_{exp})$ is a scalar, which guarantees the normalisation of $\pi(\vec{p}|\vec{\sigma}_{exp})$.
 Both the likelihood $l$ and the prior distribution $\pi_0$ are supposed to be multivariate normal distributions.
 Therefore, we can write:
 \begin{equation}
   \label{eqA2}
     \pi_0(\vec{p}) \propto \exp \left( -\frac{1}{2} (\vec{p} - \vec{p}_{ref})^T \Sigma_p^{-1} (\vec{p} - \vec{p}_{ref}) \right),
 \end{equation}
 with $\vec{p}_{ref}$ the best \apr estimate of $\vec{p}$ and $\Sigma_p$ the covariance matrix of $\vec{p}$, and:
 \begin{align}
   \label{eqA3}
     & l(\vec{\sigma}_{exp}|\vec{p}) \propto \nonumber \\
     & \exp \left(- \frac{1}{2} (\vec{\sigma}_{exp} - \mathcal{M}(\vec{p}))^T \Sigma_e^{-1} (\vec{\sigma}_{exp} - \mathcal{M}(\vec{p})) \right),
 \end{align}
 with $\Sigma_e$ the covariance matrix of $\vec{\sigma}$ and $\mathcal{M}$ the assumed to be perfect model.
 The model defects and possible biases in the experimental data are responsible for the difference between the optimal parameters and the true parameters.
 
 Since the product of two (multivariate) normal distributions is also a (multivariate) normal distribution, we also have:
 \begin{equation}
   \label{eqA4}
     \pi(\vec{p}|\vec{\sigma}_{exp}) \propto \exp \left(- \frac{1}{2} (\vec{p} - \vec{p}_{op})^T \Sigma_{op}^{-1} (\vec{p} - \vec{p}_{op}) \right),
 \end{equation}
 with $\vec{p}_{op}$ and $\Sigma_{op}$ the optimal parameter set for the model, knowing the experimental data set $\vec{\sigma}_{exp}$, and the corresponding covariance matrix, respectively.
 
 Since the GLS method requires linear relationships between variables, we need to approximate the model $\mathcal{M}$ with a Taylor series approximation:
 \begin{equation}
   \label{Taylor}
     \mathcal{M}(\vec{p}) = \mathcal{M}(\vec{p}_{ref}) + J_p (\vec{p} -\vec{p}_{ref}),
 \end{equation}
 with $J_p$ the Jacobian of the model.
 
 Therefore, we can rewrite the likelihood as:
 \begin{align}
   \label{eqA6}
     & l(\vec{\sigma}_{exp}|\vec{p}) \propto \nonumber \\
     & \exp \left(- \frac{1}{2} (\vec{\sigma}_{exp} - \mathcal{M}(\vec{p}_{ref}) - J_p (\vec{p} -\vec{p}_{ref}))^T \Sigma_e^{-1} \right. \nonumber \\
     & \qquad \left. {\color{white}\frac{1}{2}} (\vec{\sigma}_{exp} - \mathcal{M}(\vec{p}_{ref}) - J_p (\vec{p} -\vec{p}_{ref})) \right),
 \end{align}
 which can be simplified as:
 \begin{align}
   \label{eqA7}
     l(\vec{\sigma}_{exp} & |\vec{p}) \propto \nonumber \\
     & \exp \left(- \frac{1}{2} (H_{ref} - J_p \vec{p} )^T  \Sigma_e^{-1}(H_{ref} - J_p \vec{p}) \right).
 \end{align}
 with the substitution of the constant term $H_{ref}~=~\vec{\sigma}_{exp} - \mathcal{M}(\vec{p}_{ref}) + J_p \vec{p}_{ref}$.
 
 With a combination of equations \ref{eqA1}, \ref{eqA2}, \ref{eqA4}, and \ref{eqA7}, and knowing that $\pi(\vec{\sigma}_{exp})$ is a scalar, we have:
 \begin{align}
   \label{eqA8}
      \exp & \left(- \frac{1}{2} (\vec{p} - \vec{p}_{op})^T \Sigma_{op}^{-1} (\vec{p} - \vec{p}_{op}) \right) \propto \nonumber \\
      & \exp \left( -\frac{1}{2} (\vec{p} - \vec{p}_{ref})^T \Sigma_p^{-1} (\vec{p} - \vec{p}_{ref}) \right) \times \nonumber \\
      & \exp \left( - \frac{1}{2} (H_{ref} -  J_p \vec{p} )^T  \Sigma_e^{-1}(H_{ref} -  J_p \vec{p}) \right)
 \end{align}\newpage
 With this, it follows:
 \begin{align}
   \label{eqA9}
      (\vec{p} - & \vec{p}_{op})^T \Sigma_{op}^{-1} (\vec{p} - \vec{p}_{op}) + C = \nonumber \\
      & (\vec{p} - \vec{p}_{ref})^T \Sigma_p^{-1} (\vec{p} - \vec{p}_{ref}) + \nonumber \\
      & (H_{ref} - J_p \vec{p} )^T  \Sigma_e^{-1}(H_{ref} -  J_p \vec{p}),
 \end{align}
 with $C$ a constant of normalisation.
 
 Since $\vec{p}$ is the only variable in \autoref{eqA7}, the coefficients must match for the terms with $(\vec{p})^T$ on the left hand side and those with  $\vec{p}$ on the right hand side. We then have the four equations:
 \begin{align}
 \Sigma_{op}^{-1} & = \Sigma_p^{-1} + J_p^T \Sigma_e^{-1} J_p \label{e1} \\
 \Sigma_{op}^{-1} \vec{p}_{op} & = \Sigma_p^{-1}\vec{p}_{ref} + J_p^T  \Sigma_e^{-1} H_{ref} \label{e2} \\
 (\vec{p}_{op})^T \Sigma_{op}^{-1} & = (\vec{p}_{ref})^T \Sigma_p^{-1} + H_{ref}^T \Sigma_e^{-1} J_p \label{e3} \\
 (\vec{p}_{op})^T \Sigma_{op}^{-1} & \vec{p}_{op} + C = \nonumber \\
 (\vec{p}& _{ref})^T \Sigma_p^{-1} \vec{p}_{ref} + H_{ref}^T \Sigma_e^{-1} H_{ref} \label{e4}
 \end{align}
 
 Using the Woodbury matrix identity in \autoref{e1}, we have:
 \begin{equation}
   \label{eqA14}
     \Sigma_{op} = \Sigma_p - \Sigma_p J_p^T ( \Sigma_e + J_p \Sigma_p J_p^T )^{-1} J_p \Sigma_p
 \end{equation}
 
 Multiplying \autoref{e2} from the left with \autoref{eqA14}, we get:
 \strip
 \begin{align}
     \vec{p}_{op} & = \left(\Sigma_p - \Sigma_p J_p^T ( \Sigma_e + J_p \Sigma_p J_p^T )^{-1} J_p \Sigma_p \right) \left(\Sigma_p^{-1}\vec{p}_{ref} + J_p^T  \Sigma_e^{-1} H_{ref}\right)  \nonumber \\
      & = \vec{p}_{ref} + \Sigma_p J_p^T  \Sigma_e^{-1} H_{ref} - \Sigma_p J_p^T ( \Sigma_e + J_p \Sigma_p J_p^T )^{-1} J_p\vec{p}_{ref}  \nonumber \\
      &  \qquad \qquad \qquad \qquad \qquad  \qquad \qquad \qquad \qquad \qquad \qquad - \Sigma_p J_p^T ( \Sigma_e + J_p \Sigma_p J_p^T )^{-1} J_p \Sigma_p J_p^T  \Sigma_e^{-1} H_{ref}  \nonumber \\
      & = \vec{p}_{ref} - \Sigma_p J_p^T ( \Sigma_e + J_p \Sigma_p J_p^T )^{-1} J_p\vec{p}_{ref} + \Sigma_p J_p^T \left( \Sigma_e^{-1}-( \Sigma_e + J_p \Sigma_p J_p^T )^{-1}J_p\Sigma_p J_p^T  \Sigma_e^{-1} \right) H_{ref} \nonumber \\
      & = \vec{p}_{ref} - \Sigma_p J_p^T ( \Sigma_e + J_p \Sigma_p J_p^T )^{-1} J_p\vec{p}_{ref} \nonumber \\
      &  \qquad \qquad \qquad \qquad \qquad + \Sigma_p J_p^T  ( \Sigma_e + J_p \Sigma_p J_p^T )^{-1} \left( ( \Sigma_e + J_p \Sigma_p J_p^T )\Sigma_e^{-1}-J_p\Sigma_p J_p^T  \Sigma_e^{-1} \right) H_{ref}  \nonumber \\
      & = \vec{p}_{ref} - \Sigma_p J_p^T ( \Sigma_e + J_p \Sigma_p J_p^T )^{-1} J_p\vec{p}_{ref} + \Sigma_p J_p^T  ( \Sigma_e + J_p \Sigma_p J_p^T )^{-1} H_{ref} \nonumber \\
      & = \vec{p}_{ref} + \Sigma_p J_p^T ( \Sigma_e + J_p \Sigma_p J_p^T )^{-1} (H_{ref} - J_p\vec{p}_{ref})
 \end{align}
 \endstrip
 and, replacing $H_{ref}$, we finally obtain:
 \begin{align}
   \label{eqFin}
  \vec{p}_{op} & = \vec{p}_{ref} + \nonumber \\
  & \Sigma_p J_p^T ( \Sigma_e + J_p \Sigma_p J_p^T )^{-1} (\vec{\sigma}_{exp} - \mathcal{M}(\vec{p}_{ref}))
 \end{align}
 
 It is important to emphasise that equation \autoref{eqFin} is only valid as long as the hypothesis of a linear model is valid.
 However, most realistic models can be approximated by a linear model only locally.
 Therefore, $\mathcal{M}(\vec{p}_{ref})$ must be estimated reversing \autoref{Taylor}:
 \begin{equation}
   \label{Taylor2}
     \mathcal{M}_{lin}(\vec{p}_{ref}) = \mathcal{M}(\vec{p}) + J_p (\vec{p}_{ref} -\vec{p}),
 \end{equation}
 with $J_p$ the Jacobian of the model in $\vec{p}$.
 
 In order to simplify \autoref{eqFin}, we usually introduce the matrix of regression $\tilde{\Sigma}_{i}$ defined using \autoref{eq4}.
 Explicitly, the equation expands as:
 
 ~
 \vspace{-8mm}
 \begin{align}
   \tilde{\Sigma}_{i} = \mathcal{J}_i \ \Sigma \ \mathcal{J}_i^T & = \left(
      \begin{matrix}
         \Sigma_e + J_p \Sigma_p J_p^T & J_p \Sigma_p \\
         \Sigma_p J_p^T & \Sigma_p \\
      \end{matrix}
    \right) \nonumber \\
    & = \left(
      \begin{matrix}
         \tilde{\Sigma}_{DD_i} & \tilde{\Sigma}_{DI_i} \\
         \tilde{\Sigma}_{ID_i} & \tilde{\Sigma}_{II_i} \\
      \end{matrix}
    \right).
   \label{tilde}
 \end{align}
 
 We finally obtain \autoref{eq6}:
 \begin{equation}
   \vec{p}_{op} = \vec{p}_{ref} + \tilde{\Sigma}_{ID_i} \left( \tilde{\Sigma}_{DD_i} \right)^{-1} \left[ \vec{\sigma}_{exp} - \mathcal{M}_{lin}(\vec{p}_{ref})\right],
 \end{equation}
 and \autoref{eq8} follows from \autoref{eqA14}:
 \begin{equation}
   \Sigma_{op} = \tilde{\Sigma}_{II_i} - \tilde{\Sigma}_{ID_i} \left( \tilde{\Sigma}_{DD_i} \right)^{-1} \tilde{\Sigma}_{DI_i}.
 \end{equation}

 As the difference $|\vec{p}_{op} - \vec{p}|$ becomes smaller, the hypothesis of a linear model between $\vec{p}$ and $\vec{p}_{op}$ becomes more applicable, and therefore, the last two equations become more exact.
 This justifies the use of an iterative algorithm evaluating a linearisation of the model (\autoref{eq1} : $\vec{T_i}$) and its Jacobian (\autoref{eq2} : $J_{p_i}$) in  $\vec{p}_{i}$, the best evaluation of the optimal parameters currently known and then, evaluating an improved $\vec{p}_{i+1}$ from $\vec{T_i}$ and $J_{p_i}$ using \autoref{eq6}.

\section{LINP experimental data}
\label{tab_data}

In this appendix, we display the experimental data from LINP \cite{LINP} we used for the optimisation of the model parameters in \autoref{favour}. The data points has been chosen to be representative of the entire set.

To avoid overloading the computer memory, the model uncertainties are lost in the process. However, the statistics used has been chosen in order to have roughly 10\% model uncertainties for every data points independently of the absolute cross section.

\strip
\centering
\captionof{table}{List of experimental data used to adjust parameters involved in the subthreshold $K^+$ production versus the model predictions before and after optimisation.}
\begin{tabular}{|Sc|Sc|Sc|Sc|Sc|}
\hline
Kinetic & experimental & experimental & model before & model after \\
energy (MeV) & cross section (nb) & error bar (nb) & optimisation (nb) & optimisation (nb)  \\
\hline
\hline
\multicolumn{5}{|Sc|}{Pb}\\
\hline
910 & 77 & 11 & 546 & 89 \\
927 & 151 & 9 & 1110 & 116 \\
960 & 328 & 25 & 2221 & 327 \\
988 & 550 & 18 & 5036 & 787 \\
\hline
\hline
\multicolumn{5}{|Sc|}{Sn}\\
\hline
883 & 24 & 3 & 163 & 9.4 \\
910 & 75 & 7 & 300 & 43 \\
959 & 231 & 21 & 895 & 137 \\
988 & 405 & 22 & 3036 & 557 \\
\hline
\hline
\multicolumn{5}{|Sc|}{Cu}\\
\hline
840 & 8.1 & 0.9 & 9.6 & 2.3 \\
898 & 46 & 3 & 52 & 8.9 \\
927 & 81 & 5 & 229 & 30 \\
959 & 141 & 15 & 481 & 76 \\
988 & 298 & 15 & 771 & 160 \\
\hline
\hline
\multicolumn{5}{|Sc|}{C}\\
\hline
842 & 1.1 & 0.3 & 5.0 & 0.64 \\
870 & 1.8 & 0.3 & 9.3 & 2.8 \\
900 & 4.9 & 0.4 & 9.1 & 3.2 \\
905 & 6.0 & 0.5 & 16.9 & 2.4 \\
947 & 16.2 & 1.9 & 24.4 & 20.3 \\
990 & 39.0 & 2.0 & 136.1 & 18.8 \\
\hline
\hline
\multicolumn{5}{|Sc|}{Be}\\
\hline
835 & 0.25 & 0.21 & 4.55 & 1.69 \\
878 & 1.47 & 0.28 & 15.48 & 3.17 \\
918 & 3.9 & 0.4 & 27.5 & 6.3 \\
960 & 9.2 & 1.1 & 68.8 & 14.8 \\
990 & 21 & 1.0 & 130 & 27 \\
\hline
\end{tabular}

\endstrip

\end{appendices}
\newpage~\newpage~

\bibliographystyle{unsrt} 
\bibliography{paper} 

\end{document}